\documentclass[12pt,aps,nofootinbib]{revtex4}
\usepackage{graphicx}
\usepackage{amssymb}
\usepackage{amsmath}
\usepackage{bm}
\usepackage{times}
\usepackage{epsfig}
\usepackage{epstopdf}
\usepackage{color}
\usepackage{multirow}
\usepackage[colorlinks=true,linkcolor=blue,citecolor=blue]{hyperref}

\begin{document}
\title{Investigating the Effect of Family Non-universal $Z^\prime$ Boson in $B \to \phi\phi$ Decay}
\author{Ying Li\footnote{Email:liying@ytu.edu.cn}}
\affiliation{Department of Physics, Yantai University, Yantai
264005, China}
\begin{abstract}
Within the perturbative QCD approach, we re-calculate the branching
ratio and polarization fractions of the pure annihilation decay $B
\to \phi \phi$ in both the standard model (SM) and the family
non-universal $Z^\prime$ model. We find that this decay is dominated
by the longitudinal part, while the transverse parts are negligibly
due to the absence of the $(S-P)(S+P)$-type operator. In SM, the
branching ratio is predicted  as $(4.4^{+0.8+0.3}_{-0.6-0.5}) \times
10^{-8}$, which is larger than the previous predictions. With an
additional $Z^\prime$ boson, the branching ratio can be enhanced by
a factor of 2, or reduced  one half in the allowed parameters space.
These results will  be tested by the ongoing LHCb experiment and
forthcoming Super-B experiments. Moreover, if the $Z^\prime$ boson
could be directly detected at hadron collider, this decay can be
used to constrain its mass and the couplings in turn.
\end{abstract}
\date{\today}
\maketitle
\section{Introduction}
Despite the fact that the standard model of particle physics has
various predictions that are in accordance with  experimental data,
it is generally viewed as an effective realization of an underlying
theory to be discovered yet. Interestingly to understand the
hierarchy problem of the Higgs mass, neutrino masses, and  the CP
asymmetry,  one is often allured to resort to the new physics (NP)
beyond SM. If existing, the NP degree of freedom may manifest
itself either directly at the hadron collider or indirectly at low
energy via its effects to observables that have been precisely
constrained. Over the past years, processes induced by
flavor-changing-neutral-current (FCNC) have been under sharper
scrutiny, as these processes are forbidden at the tree level and
thus arise only at the loop level within SM. Many NP models have
different patterns with SM and enhance the FCNC transition at the
tree or loop level, which are likely to affect some physical
observables sizably compared to SM.

The rare decay $B \to \phi\phi$ is of this type and proceeds via a
FCNC process $b\to d \bar s s$. Moreover, since all quarks in the
final states are different from those in the initial $B$ meson, this
decay involves only the pure annihilation topology. As a consequence
of  the power counting rules derived from the heavy quark effective
theory,  its branching ratio is expected to be very tiny. Meanwhile,
on the experimental side, the signature of this decay is very clean.
Due to these advantages,  the $B \to \phi\phi$ has thus received
considerable attentions in both theoretical \cite{BarShalom:2002sv,
Lu:2005be, Beneke:2006hg} and experimental sides
\cite{Aubert:2008fq, Olaiya:2008zh} in the past few years.

To the  best of our knowledge, an annihilation amplitude involving
two light mesons suffers from the endpoint divergence, and many
approaches have been advocated for dealing with it. In
\cite{BarShalom:2002sv}, by introducing the effective gluon mass
$m_g=500\pm 200 \mathrm{MeV}$, the authors predicted $Br(B^0 \to
\phi \phi)= (2.1^{+1.6}_{-0.3})\times 10^{-9}$ in SM. While in the
R-parity violating supersymmetric model, the branching fraction of
this decay could be enhanced to $10^{-7}$. In the QCD factorization
approach, the endpoint singularity has been usually parameterized by
two free parameters $\rho_A$ and $\phi_A$ in a phenomenology way,
which are mode-dependent and cannot be calculated directly. As a
result, only the upper limit of this decay $10^{-8}$ has been
presented in \cite{Beneke:2006hg}. By keeping the intrinsic
transverse momenta $\mathrm{k}_T$ of the valence quarks in the
perturbative QCD (PQCD) approach, the annihilation topologies could
be calculated directly, as the divergence can be eliminated by the
Sudakov form factor and the threshold resummation. Within the PQCD
approach, its branching ratio has been predicted to be
$(1.89^{+0.61}_{-0.21}) \times 10^{-8}$ in \cite{Lu:2005be}, in
which the longitudinal polarization fraction was estimated to be
about $65\%$. However, as discussion the decay modes $B \to \phi
K^{*}$ \cite{Chen:2002pz, Li:2004ti, Li:2004mp}, it has been known
that the longitudinal polarization fraction about $48\%$ was
measured in experiments. In the PQCD framework, the annihilation
contribution from the $(S-P)(S+P)$ operators enhances the amplitudes
remarkably due to the helicity flip, so the so-called "polarization
anomaly" could be well understood. However, because the $(S-P)(S+P)$
operator vanishes in this mode, it is hard for us to understand the
large transverse polarization $35\%$ predicted  in \cite{Lu:2005be}.
Therefore, it is necessary to re-analyse this decay in SM within the
PQCD approach.

As stated above, in SM, the decay $B \to \phi\phi$ is expected to
have a small branching ratio, which allows us to search for possible
NP effects. Hence, another purpose of this work is to explore the
effects of an extra $Z^\prime$ boson on this decay, which is allowed
in a few well motivated extensions of SM due to an additional
$\mathrm{U}(1)^\prime$ gauge symmetry. Among many $Z^\prime$ models,
the most general one is the family non-universal $Z^\prime$ model,
which can be realized in various grand unified theories,
string-inspired models, dynamical symmetry breaking models, and the
little Higgs models, just to name a few \cite{Zprime}. The
$Z^\prime$ boson in different representative models has been
directly searched at colliders as well as indirectly probed via a
variety of precision data \cite{Hayden:2013sra}, which put limits on
its gauge coupling and/or mass. In such a model, the nonuniversal
$Z^\prime$ couplings to fermions could lead to FCNC at the tree
level, which may enhance the branching ratios of some rare $B$
decays dominated by penguin operators. In recent years, the effects
of the $Z^\prime$ boson have been studied extensively in the low
energy flavor physics phenomenology, such as in $B$ physics, top
physics, and lepton decays \cite{Chiang:2013aha}.

In this work, we will first reanalyze $B \to \phi\phi$ in SM within
the PQCD approach in Sec.\ref{sec:2}, and find that the results for
branching ratios are larger than the predictions in
\cite{Lu:2005be}. We then in Sec.\ref{sec:3} consider the
contribution of the non-universal $Z^\prime$ boson, which could
change the branching ratio in the suitable parameters space. At
last, we summarize this work in Sec. \ref{sec:4}.

\section{Calculation in SM}\label{sec:2}
In SM, the relevant effective weak Hamiltonian related to $B \to \phi\phi$ is given by:
\begin{eqnarray}\label{hamiton}
{\cal H}_{eff}^{SM}={G_F\over \sqrt 2}V^*_{tb}V_{td}\sum\limits_{i=3}^{10} C_iO_i.
\end{eqnarray}
$O_i$ are the four-quark operators and $C_i$ are the corresponding Wilson coefficients, whose explicit expressions are refereed to \cite{Lu:2005be}. $V_{tb}$ and $V_{td}$ are the Cabibbo-Kabayashi-Maskawa (CKM)  matrix elements. Then, the decay width for this decay is written as
\begin{eqnarray}
\Gamma=\frac{p_c}{8S\pi m_B^2}\sum\limits_{\sigma =L,T}{\cal M}^{\sigma\dag}{\cal M}^{\sigma} ,
\end{eqnarray}
where $p_c$ is the momentum of the outgoing mesons and $S=2$ comes from the identical final state particles. The decay amplitude ${\cal M}^{\sigma}$ will be calculated later, where the subscript $\sigma$ denotes the helicity states of the two vector mesons with $L(T)$ standing for the longitudinal (transverse) component. Furthermore, the amplitude ${\cal M}^{\sigma}$ can be decomposed into:
\begin{eqnarray}
{\cal M}^{\sigma}=m_B^2{\cal M}_L+m_B^2{\cal M}_N\epsilon^{\ast}_2(\sigma=T)\cdot \epsilon^{\ast}_3(\sigma=T) 
+i{\cal M}_T\epsilon_{\mu\nu\rho\sigma}\epsilon_2^{\mu\ast} \epsilon_3^{\nu\ast}P_2^{\rho}P_3^{\sigma},
\end{eqnarray}
where $\epsilon_{2(3)}$ and $P_{2(3)}$ are the polarization vector and the  four-momentum of the final state vector meson, respectively. Conventionally, the longitudinal $H_{00}$ helicity amplitudes and the transverse helicity amplitudes $H_{\pm\pm}$ are defined by
\begin{eqnarray}
H_{00}&=&m^2_B{\cal M}_L\\
H_{\pm\pm}&=&m^2_B{\cal M}_N\mp m_{\phi}^2\sqrt{\kappa^{2}-1}{\cal M}_T,
\end{eqnarray}
with the helicity summation,
\begin{equation}
\sum\limits_{\sigma =L,T}{\cal M}^{\sigma\dag}{\cal M}^{\sigma}=|H_{00}|^2+|H_{++}|^2+|H_{--}|^2,
\end{equation}
and $\kappa=(P_2\cdot P_3)/ {m_{\phi}^2}$. Another equivalent set of  definitions of helicity amplitudes is also used,
\begin{eqnarray}
&& A_0=-\zeta m^2_B{\cal M}_L,\nonumber\\
&& A_{\|}=\zeta \sqrt{2} m^2_B{\cal M}_N, \nonumber \\
&& A_{\perp}=\zeta m_{\phi}^2\sqrt{\kappa^{2}-1}{\cal M}_T,
\end{eqnarray}
with the normalization factor $\zeta$ satisfying
\begin{eqnarray}
 |A_0|^2+|A_{\|}|^2+ |A_{\perp}|^2= 1,
\end{eqnarray}
where the notations $A_0$, $A_{\|}$, $A_{\perp}$ denote the longitudinal ($f_L$),  parallel ($f_{\|}$),  and perpendicular ($f_{\perp}$) polarization fractions, respectively.

\begin{figure}[ht]
\centering
\includegraphics[width=0.4\textwidth]{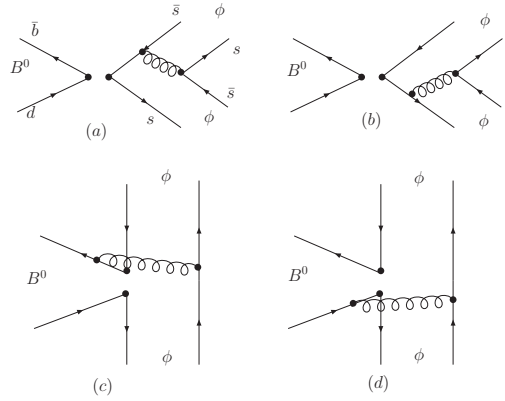}
\caption{Feynman diagrams for the $B^0\rightarrow \phi\phi$ decay in
the PQCD approach. }\label{fig:1}
\end{figure}

Now, we will evaluate the hadronic matrix elements ${\cal M}_L$, ${\cal M}_N$  and ${\cal M}_T$ using the PQCD approach. According to the effective Hamiltonian in Eq.~(\ref{hamiton}), we draw the lowest order diagrams of $B \to \phi\phi$ as shown in Fig.\ref{fig:1}. In PQCD, the decay amplitude is factorized into the soft part $\Phi$, the hard part $H$, and the harder one $C_i$ characterized by different scales. It is conceptually written as,
\begin{eqnarray}
 {\cal M}&\sim &\int dx_1dx_2dx_3b_1db_1b_2db_2b_3db_3Tr[C(t)\Phi_B(x_1,b_1)\nonumber \\
&&\Phi_{\phi}(x_2,b_2) \Phi_{\phi}(x_3,b_3)H(x_i,b_i,t)S_t(x_i)e^{-S(t)}],
\end{eqnarray}
where $x_i$ denotes the momentum fraction of a light quark in each meson, and $b_i$ is the conjugate space coordinate of the transverse momentum. $Tr$ means the trace over Dirac and color indices, and $C(t)$ is the Wilson coefficient evaluated at scale $t$. The universal wave function $\Phi_M(M=B,\phi)$ describes hadronization of a quark and an anti-quark into the meson $M$, whose structure can be found in \cite{Lu:2005be, Chen:2002pz, Ali:2007ff}. $H$ is the six-quark hard scattering kernel, which consists of the effective four quark operators and a hard gluon attaching to the spectator quark in the decay, so it can be perturbatively calculated. The function $S_t(x_i)$ describes the threshold resummation which smears the end-point singularities. The last term $e^{-S(t)}$, coming from the resummation of the double logarithm $\ln^2 k_T$, is the Sudakov form factor which suppresses soft dynamics effectively.

As shown in Fig.\ref{fig:1}, there are four kinds of Feynman diagrams  contributing to the $B \to \phi \phi$ decay at leading order. They involve two types: factorizable diagrams ($a$) and ($b$), and non-factorizable diagrams ($c$) and ($d$). After calculating these diagrams, we can get the amplitudes as follows:
\begin{eqnarray}
 {\cal M}_{i=L,N,T}& =&\frac{2G_F}{\sqrt{2}} V_{tb}V_{td}^{*}\Bigg\{f_{B}F_{ann}^{LL,i}\left[C_3+\frac{1}{3}C_{4}-\frac{1}{2}C_9  -\frac{1}{6}C_{10}\right]+M_{ann}^{LL,i}\left[C_{4}-\frac{1}{2}C_{10}\right]\nonumber \\&+& f_{B} F_{ann}^{LR,i}\left[C_5+\frac{1}{3}C_{6}-\frac{1}{2}C_7-\frac{1}{6}C_{8} \right]
  +M_{ann}^{SP,i} \left[C_6-\frac{1}{2}C_{8}\right]\Bigg\} .
\end{eqnarray}
where $i=L,N,T$ stands for the longitudinal polarization and the two transverse polarizations.  $f_{B} F_{ann}^{LL(LR)}$ comes from the contribution of the factorizable diagrams with the operators $(V-A)(V-A)$ or $(V-A)(V+A)$, and $f_{B}$ is the decay constant of the $B$ meson. $M_{ann}^{LL(SP),i}$ is the non-factorizable amplitude with the operator $(V-A)(V-A)$ or $(S-P)(S+P)$, and the latter operator is from the Fierz transformation of the operator $(V-A)(V+A)$. In \cite{Ali:2007ff,Chen:2002pz}, the authors had listed all formulae of $f_{B} F_{ann}^{LL(LR),i}$ and $M_{ann}^{LL(SP),i}$ at leading order in detail, thus it is not necessary to duplicate them in the current work.


Due to the current conservation, for the longitudinal and parallel polarization parts,  the contributions from the factorizable diagrams $(a)$ and $(b)$  are canceled exactly by each other, leading to $f_B F_{ann}^{LL(LR),L(N)}=0$. Therefore, there is only $f_BF_{ann}^{LL(LR),T}$ left for the factorizable diagrams, but it is suppressed by $(m_\phi/m_B)^2$. For the non-factorizable diagrams $(c)$ and $(d)$, the longitudinal parts give the leading and dominant contribution, and other terms are suppressed by $(m_\phi/m_B)^2$. That is to say, the contribution of the longitudinal and parallel polarization is only from the non-factorizable diagrams, but the latter one is suppressed by $4\%$. Although the perpendicular part receives another effect from the diagrams $(a)$ and $(b)$, but their contribution is negligible. Thus, the transverse parts can be dropped safely in SM.

In the numerical calculation, we must input the $B$ and $\phi$ meson distribution amplitudes, which are nonperturbative parameters. For the $B$ meson, we employ the function
\begin{eqnarray}
\phi_{B}(x,b)=N_{B}x^{2}(1-x)^{2}\exp \left[ -\frac{1}{2} \left( \frac{xm_{B}}{\omega _{B}}\right) ^{2} -\frac{\omega_{B}^{2}b^{2}}{2}\right]\label{bw} \;,
\end{eqnarray}
where the shape parameter $\omega_{B}=0.4$ GeV has been adopted in all  previous analysis of exclusive $B$ meson decays. The normalization constant $N_{B}= 91.784$ GeV is related to the decay constant $f_{B}=190$ MeV. Since the $\phi$ meson is a vector particle, there are six distribution amplitudes up to twist 3, and all of them have been calculated in QCD sum rules \cite{Ball:1998sk}. The formulae have been also given explicitly in \cite{Chen:2002pz,Ball:1998sk}.

Honestly speaking, there are many theoretical uncertainties in our calculation. For the penguin-dominated decays, one of the important uncertainties is from the hard scales $t$, which are defined as the invariant masses of internal particles and are required to be higher than the factorization scale $1/b$, $b$ being the transverse extents of the mesons. Another large uncertainty comes from the distribution amplitude of $B$ meson, since it cannot be calculated directly from the first principle. Varying the hard scales $t$ between $0.75-1.25$ times the center values and the shape parameter $\omega_{B}=0.40\pm 0.05$, we then obtain the $B \to \phi \phi$ branching ratio
\begin{eqnarray}
Br(B^0\rightarrow\phi\phi)=(4.4^{+0.8+0.3}_{-0.6-0.5})\times 10^{-8}.
\end{eqnarray}
The uncertainties from the $\phi$ meson distribution amplitudes are less than $20\%$, so we will not discuss them here. The above branching ratio can be measured at the Large Hadron Collider beauty (LHCb) experiments or the Super-$B$ factory in future, which helps test SM. For the longitudinal polarization fraction, it is given by
\begin{eqnarray}
f_L\approx 1.
\end{eqnarray}
Compared with the results of \cite{Lu:2005be}, our branching ratio
is about twice larger than theirs. For the polarization, it does not
agree with theirs either.  Furthermore, the large longitudinal
polarization fraction in the annihilation decay mode $B^0 \to K^{*+}
K^{*-} $ has been also confirmed in \cite{Zhu:2005rt}.  With the
formulae and parameters given in \cite{Lu:2005be}, we get the
branching ratio $3.9\times 10^{-8}$ and $f_L\approx 1$, which agree
with present results considering the difference of Wilson
coefficients and other parameters.

\section{Effect of $Z^\prime$ Boson}\label{sec:3}
Now we are in position to analyze this process with an extra gauge
boson $Z^\prime$.  In the gauge basis, ignoring the mixing between
$Z$ and $Z^\prime$, we write the $Z^\prime$ term of the
neutral-current Lagrangian as
\begin{eqnarray}
{\cal L}^{Z^\prime} =-g^\prime Z^{\prime {\mu}}\sum_{i,j} {\overline \psi_i^I} \gamma_{\mu}
  \left[ (\epsilon_{\psi_L})_{ij} P_L + (\epsilon_{\psi_R})_{ij} P_R \right] \psi^I_j ~,
\end{eqnarray}
where $i$ is the family index and labels the fermions. $g^\prime$ is the gauge  coupling constant at the electro-weak scale $M_W$, and $P_{L,R}=(1\mp\gamma_5)/2$. The superscript $I$ refers to the gauge interaction eigenstate, and $\epsilon_{\psi_L}$ ($\epsilon_{\psi_R}$) denotes the left-handed (right-handed) chiral coupling. After rotating to the physical basis, the fermion Yukawa coupling matrices $Y_{\psi}$ in the weak basis can be diagonalized as
\begin{eqnarray}
 Y_{\psi}^D = V_{\psi_R} Y_{\psi} V_{\psi_L}^\dagger
\end{eqnarray}
using the unitary matrices $V_{\psi_{L,R}}$ in $\psi_{L,R}=V_{\psi_{L,R}}  \psi_{L,R}^I$, where $\psi_{L,R}^I \equiv P_{L,R} \psi^I$ and $\psi_{L,R}$ are the mass eigenstate fields. The CKM matrix is usually given by
\begin{eqnarray}
 V_{\rm CKM} = V_{u_L} V_{d_L}^{\dagger}.
\end{eqnarray}
So, the chiral $Z^\prime$ coupling matrices in the mass basis of
down-type  quarks could be written as
\begin{eqnarray}
B^L_d&\equiv& V_{d_L} \epsilon_{d_L} V_{d_L}^{\dagger} ~, \\
B^R_d&\equiv& V_{d_R} \epsilon_{d_R} V_{d_R}^{\dagger}.
\end{eqnarray}
If $\epsilon_{d_{L,R}}$ are not proportional to the identity matrix,
$B^{L,R}_d$  will have nonzero off-diagonal elements that induce
FCNC interactions. In the current work, we will assume that the
right-handed couplings are flavor-diagonal for simplicity.

With nonzero flavor-diagonal matrix elements, the $Z^\prime$ boson
contributes  to FCNC at the tree level, and its contribution will
interfere with SM contributions. In particular, the flavor-changing
couplings of the $Z^\prime$ boson with the left-handed fermions will
contribute to the $O_9$ and $O_7$ operators for the left
(right)-handed couplings at the flavor-conserving vertex, i.e.,
$C_{9,7}(M_W)$ receive contributions from the new $Z^\prime$ boson.
Then, the $Z^{\prime}$ part of the effective Hamiltonian for $b\to
d\bar{s}s$ transitions has the form
\begin{equation}\label{heffz1}
 {\cal H}_{eff}^{\rm Z^{\prime}}=- \frac{4 G_F}{\sqrt{2}}\left(\frac{g^\prime M_Z}{g_Y M_{Z^\prime}}\right)^2 B^L_{db}
 \left( B^L_{ss} O_9 + B^R_{ss} O_{7} \right)+ \mbox{h.c.} ~,
\end{equation}
where $g_Y=e/(\sin{\theta_W}\cos{\theta_W})$ and $M_{Z^{\prime}}$ is
the  mass of the new gauge boson. $O_{7,9}$ are the effective
operators in SM. Due to the hermiticity of the effective
Hamiltonian, we always assume that the diagonal elements of the
effective coupling matrices $B_{qq}^{L,R}$ are real. However, there
is still a new weak phase $\phi$ in the off-diagonal one of
$B_{bd}^{L}$. Compared with Eq.(\ref{hamiton}), the resultant
$Z^\prime$ contributions to the Wilson coefficients are
\begin{eqnarray}
\Delta  C_{9,7}=4\frac{|V_{tb}V_{td}^{\ast}|}{V_{tb}V_{td}^{\ast}}\xi^{L,R}e^{-i\phi},
\end{eqnarray}
with
\begin{eqnarray}
\xi^{L,R}=\left(\frac{g^{\prime}M_Z}
{g_YM_{Z^{\prime}}}\right)^2\left|\frac{B_{db}^LB_{ss}^{L,R}}{V_{tb}V_{td}^{\ast}}
\right|.
\end{eqnarray}
Since the heavy degrees of freedom in the theory have already been
integrated out  at the scale $M_W$, the RG evolution of the Wilson
coefficients after including the new contributions from $Z^\prime$
is exactly the same as in SM \cite{Buchalla:1995vs}.

Generally, we always suppose $g^\prime\approx g_Y$ if both the
$U(1)$ and $U^\prime(1)$  gauge groups have the same origin from
some grand unified theories. Though the $Z^\prime$ boson has not
been detected in Large Hadron Collider (LHC) experiments, we always
expect the mass $M_{Z^\prime}$ to be at the TeV scale, which would
lead to $M_Z/M_{Z'} \approx 0.1$. In order to explain the mass
differences of $B_q - \overline B_q\,(q=d,s)$ and the $CP$ asymmetry
anomalies in $B \to \phi K, \pi K $, $|B_{qq}^{L,R}| $ should be of
${\cal O}(1)$. More about constraints on these parameters are
refereed to \cite{Buras:2012jb, chweichiang,changqin}. To quantify
the effects of the $Z^\prime$ boson, we consider $\xi^{L,R} \in
[0.001,0.02]$ in the following discussion. Moreover, for the new
weak phase $\phi$, we treat it as a free parameter.

Our analyses are divided into the following three scenarios with different simplifications, namely,
\begin{itemize}
  \item S1: Ignoring the right-hand couplings, i.e., $\xi^{R}=0$.
  \item S2: Supposing that the left-hand couplings share the same values a thse right-hand values, i.e., $\xi^{L}=\xi^{R}$.
  \item S3: Allowing arbitrary values for $\xi^{L,R}$ without any simplifications.
\end{itemize}
With the possible parameter space, we evaluate the $B \to \phi \phi$
branching ratios under the different scenarios together with the SM
contribution as
\begin{eqnarray}\label{results}
\mathrm{Br}(B \to\phi \phi)=\left\{
\begin{array}{ll}
(3.6^{+0.5+0.3+2.8}_{-0.5-0.4-0.8})\times 10^{-8}, & \hbox{S1;} \\
(5.1^{+0.9+0.5+0.8}_{-0.7-0.5-2.0})\times 10^{-8}, & \hbox{S2;} \\
(5.1^{+0.9+0.5+2.9}_{-0.7-0.5-3.2})\times 10^{-8}, & \hbox{S3;} \\
(4.4^{+0.8+0.3}_{-0.6-0.5})\times 10^{-8}, & \hbox{SM,}
\end{array}
\right.
\end{eqnarray}
where the first two errors are from uncertainties of PQCD, i.e. the
shape  parameter $\omega_B$ and the hard scale $t$.  For the
$Z^\prime$ contribution, we scan all possible parameter space
($\xi^{L,R}$ and the new weak phase $\phi$), and get the third
uncertainties. As for the center values, we take $\xi^{L,R} =0.01$
and $\phi=0$. Under S1, it is clear that the $Z^\prime$ boson plays
a destructive role for the branching ratio, while the branching
ratio will be enhanced after adding the contribution from the
right-hand couplings under S2 and S3. Since only one strong phase
exists, there is no $CP$ asymmetry in this decay. The polarizations
are almost unchanged, though the new $Z^\prime$ particle could
change the transverse parts of the amplitudes.
\begin{figure}[ht]
\centering
\includegraphics[width=0.4\textwidth]{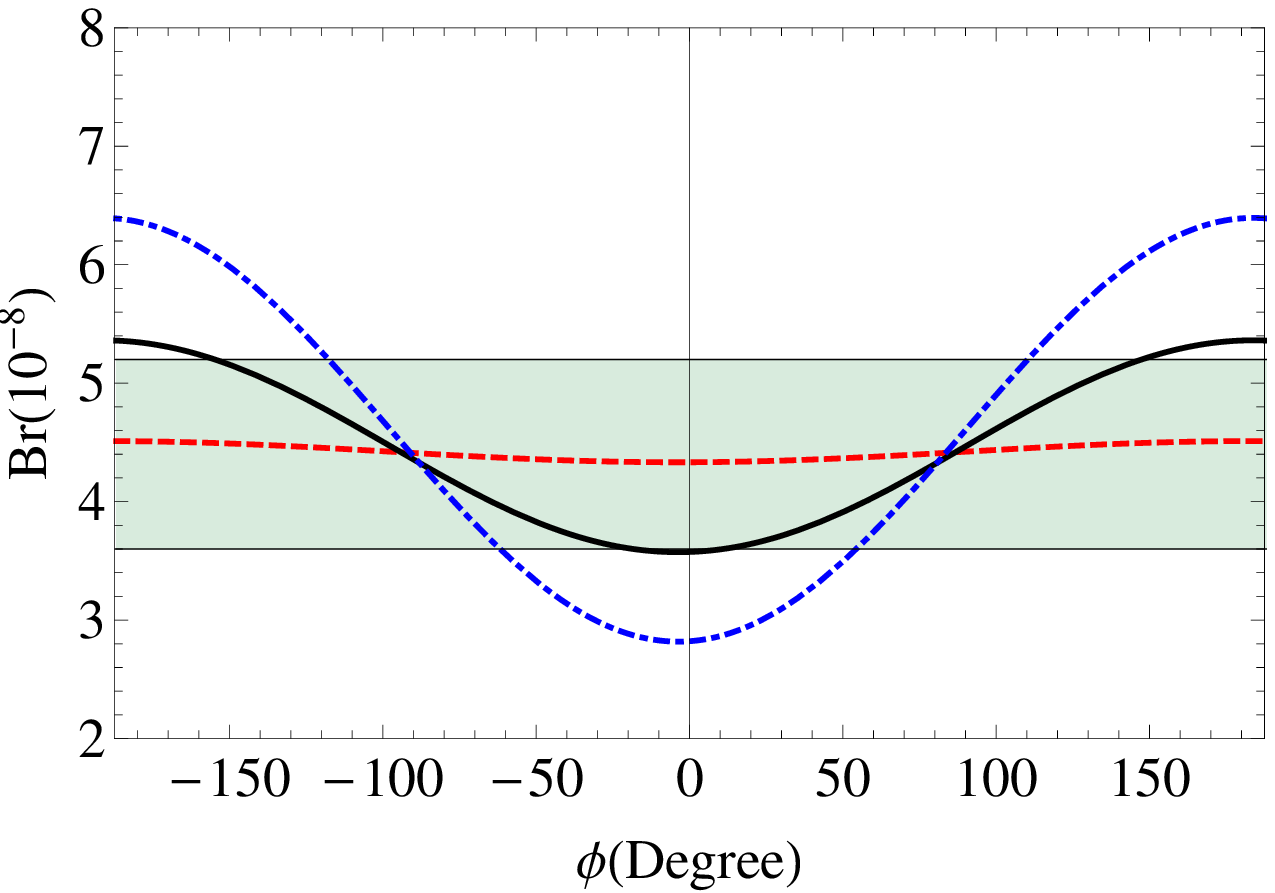}
\includegraphics[width=0.4\textwidth]{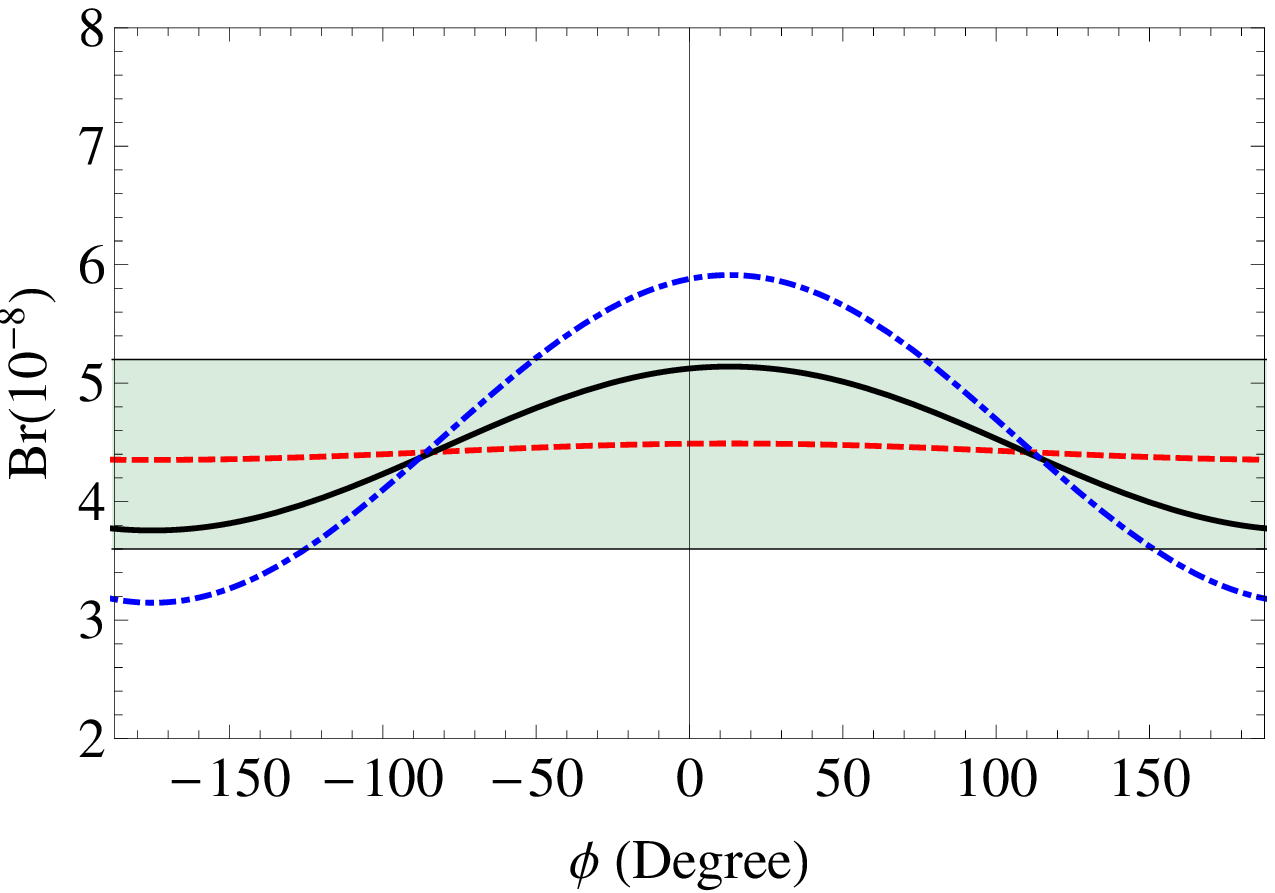}
\caption{Variation of the branching ratio with the new weak phase $\phi$ under S1(left panel) and S2 (right panel), where the dashed (red), solid (black) and dot-dashed (blue) lines correspond to $\xi=0.001$, $0.01$ and $0.02$, respectively. The range with horizontal lines shows the prediction in SM after adding the two errors in quadrature.}\label{fig:2}
\end{figure}
\begin{figure}[ht]
\centering
\includegraphics[width=0.4\textwidth]{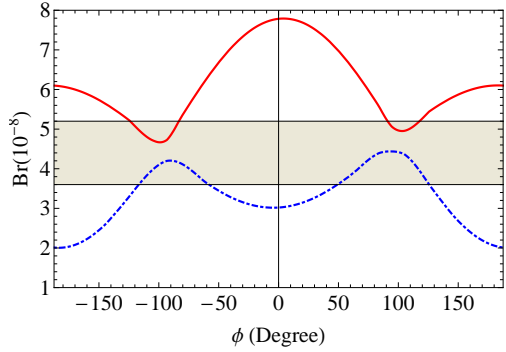}
\caption{Variation of the branching ratio with the new weak phase  $\phi$ under S3, where the solid (red) and dot-dashed (blue) curves correspond to maximal and minimal values, respectively. The range with horizontal lines shows the prediction in SM after adding the two errors in quadrature.}\label{fig:3}
\end{figure}

To study the effect of the $Z^\prime$ boson clearly, we plot the
variation of the branching ratio as a function of the new weak phase
$\phi$ with different values of $\xi=0.001,0.01,0.02$ under S1 (left
panel) and S2 (right panel), as shown in Fig. \ref{fig:2}. According
to these plots, we note that if $\xi \leq 0.001$, i.e. a heavy
$Z^\prime$ boson, the new physics effect is too small to be
detected. Even for $\xi \approx  0.01$ under both scenarios, its
effect is also hard to measure in experiments, because it will be
buried by the uncertainties of PQCD in SM. While $\xi \approx 0.02$,
the $Z^\prime$ boson will change the branching ratio remarkably, but
the trends are different for different scenarios. Under S1, the
branching ratio becomes larger and exceeds the predicted range in SM
with a large phase, while it becomes smaller with a small weak
phase, which can be seen from the left panel of Fig.\ref{fig:2}. For
S2, as seen from the right panel, it has an opposite situation. As
for S3, by varying $\xi^{L}$ and $\xi^{R}$ independently, we present
the maximal and minimal curves of the branching ratio as functions
of $\phi$ in Fig. \ref{fig:3}. It is found that the range of the
branching ratio is much larger than the SM predictions, which is
also shown in Eq.~(\ref {results}). When $\phi=0$, the maximal value
is about $10 \time 10^{-8}$, which is about twice of prediction of
SM. On the contrary, by setting $\phi=\pm 180^\circ$, it will be
decreased to half of the center value of the SM prediction.  All the
above results can be tested in the current LHCb experiments or at
the Super-$B$ factory in future. Moreover, if the $Z^\prime$ boson
would be detected in future, the observation of this mode will in
turn help us constrain the $Z^\prime$ mass and its couplings to
fermions.

\section{Summary}\label{sec:4}
In this work, we have re-calculated the branching ratio and the
polarization fractions of the pure annihilation decay $B \to \phi
\phi$ within the perturbative QCD approach in both SM and the
non-universal $Z^\prime$ model. We found that this mode is
longitudinal part dominated and its longitudinal polarization
fraction is about 1 because of absence of contributions from the
operator $(S-P)(S+P)$. The branching ratio is estimated to be
$(4.4^{+0.8+0.3}_{-0.6-0.5})\times 10^{-8}$, which may be measured
in the ongoing LHCb experiments or at the Super-$B$ factory in
future. Considering the effect of an additional $Z^\prime$ boson, we
found that the branching ratio may be enhanced by a factor of 2, or
reduced to half in the allowed parameter space, as shown in
Fig.\ref{fig:3}. Thus, if this mode could be measured in the LHCb
experiments and/or at the Super-$B$ factory, it will provide a test
of SM and the non-universal $Z^\prime$ model. Furthermore, if the
$Z^\prime$ boson could be detected, these results can be used to
constrain its mass and couplings in turn.

\section*{Acknowledgments}
Y.Li thanks Cai-Dian Lu, Y.-L. Shen and Z.-T. Zou for valuable discussions and comments.
This work is supported by the National Science Foundation (nos. 11175151 and 11235005),
the Natural Science Foundation of Shandong Province (ZR2010AM036) and the Program
for New Century Excellent Talents in University (NCET) by Ministry of Education of P.R. China.

\end{document}